\documentclass[a4paper, traditabstract, letter]{aa} % for the abstract without structuration 
\pdfoutput=1
\usepackage{graphicx}
\usepackage{txfonts}
\usepackage[breaklinks, colorlinks, citecolor=Cerulean]{hyperref}
\usepackage{url}
\usepackage{natbib}
\usepackage{amssymb}
\usepackage{booktabs}
\usepackage{subfig}
\usepackage{commath}
\usepackage{microtype}

\newfont{\gwpfont}{cmssq8 scaled 1000}
\newcommand{\rexcess}{{\gwpfont REXCESS}}
\def\xmm{XMM-{\it Newton}}
\def\planck{{\it Planck}}

\def\M500{M_{500}}
\def\R500{R_{500}}

\def\Mv {M_{\rm 500}}
\def \Rv {R_{500}}

\def\lesssim{\mathrel{\hbox{\rlap{\hbox{\lower4pt\hbox{$\sim$}}}\hbox{$<$}}}}
\def\gtrsim{\mathrel{\hbox{\rlap{\hbox{\lower4pt\hbox{$\sim$}}}\hbox{$>$}}}}

\newcommand{\propsim}{\lower 3pt \hbox{$\, \buildrel {\textstyle
       \propto}\over {\textstyle \sim}\,$}}

\begin{document}
\title{The hot gas content of fossil galaxy clusters}   
\author{G.W. Pratt\inst{1}, E. Pointecouteau\inst{2,3}, M. Arnaud\inst{1} and R.F.J. van~der~Burg\inst{1}
}
\authorrunning{G.W. Pratt et al.}
\titlerunning{The hot gas content of fossil galaxy clusters}
 \institute{
 $^1$ Laboratoire AIM, IRFU/Service d'Astrophysique - CEA/DRF - CNRS - Universit\'{e} Paris Diderot, B\^{a}t. 709, CEA-Saclay, F-91191 Gif-sur-Yvette Cedex, France \\
 \email{gabriel.pratt@cea.fr} \\ 
 $^2$ CNRS, IRAP, 9 Av. colonel Roche, BP 44346, F-31028 Toulouse Cedex 4, France\\
 $^3$ Universit\'{e} de Toulouse, UPS-OMP, IRAP, F-31028 Toulouse Cedex 4, France%\\
}
\date{Received 9 March 2016; accepted 23 April 2016}
\abstract{We investigate the properties of the hot gas in four fossil galaxy systems detected at high significance in the \planck\ Sunyaev--Zeldovich (SZ) survey.  \xmm\ observations reveal overall temperatures of $kT \sim 5-6$ keV and yield hydrostatic masses $M_{\rm 500, HE} \gtrsim 3.5 \times 10^{14}$ M$_{\odot}$, confirming their nature as bona fide massive clusters. We measure the thermodynamic properties of the hot gas in X-rays (out to beyond $\Rv$ in three cases) and derive their individual pressure profiles out to $R \sim 2.5\, R_{500}$ with the SZ data. We combine the X-ray and SZ data to measure hydrostatic mass profiles and to examine the hot gas content and its radial distribution. The average Navarro--Frenk--White (NFW) concentration parameter, $\langle c_{500} \rangle = 3.2\pm0.4$, is the same as that of relaxed `normal' clusters.  The gas mass fraction profiles exhibit striking variation in the inner regions, but converge to approximately the cosmic baryon fraction (corrected for depletion) at $\Rv$. Beyond $\Rv$ the gas mass fraction profiles again diverge, which we interpret as being due to a difference in gas clumping and/or a breakdown of hydrostatic equilibrium in the external regions. Our observations point to considerable radial variation in the hot gas content and in  the gas clumping and/or hydrostatic equilibrium properties in these fossil clusters, at odds with the interpretation of their being old, evolved, and undisturbed. At least some fossil objects appear to be dynamically young.}

   \keywords{galaxies: clusters: general -- galaxies: clusters: intracluster medium  -- X-rays: galaxies: clusters}

   \maketitle
%
%________________________________________________________________

\section{Introduction}\label{sec:intro}
%
%%%%%%%%%%%%%%%%%%%%%%%%%%%%%%%%%%%%%

When \citet{pon94} first looked at the X-ray source RX\,J1340.6+4018, they found an apparently isolated elliptical galaxy with an extended group-scale X-ray emission. To explain its unusual characteristics, they suggested that it was the result of the coalescence of a group of galaxies into a single giant elliptical, in the process retaining the X-ray emission characteristic of the original halo. Such objects would have formed long ago, leading \citeauthor{pon94} to coin the term `fossil galaxy group' to describe them. 

\citet{jon03} defined a fossil group as having a magnitude gap of $\Delta\, m_{12} \geq 2$ in the $R$ band between the brightest and the second brightest galaxy within $0.5\, R_{200}$\footnote{$R_{\delta}$ is the radius within which the total density is $\delta$ times the critical density at the redshift of the object.}, and an X-ray luminosity of $L_{\rm X, bol} \geq 5 \times 10^{41} h_{70}^{-2}$ erg s$^{-1}$. The X-ray criterion ensures a group-scale halo, while the optical criterion ensures that there are no $L^*$ galaxies inside the radius for orbital decay by dynamical friction. Other authors have modified the optical criterion; for instance, \citet{voe10} use a criterion of $\Delta\, m_{12} \geq 1.7$. We use the \citeauthor{voe10} definition here.

%-----------------------------Figure Start------------------------------
\begin{figure*}[!ht]
\begin{center}
\includegraphics[bb=90 360 510 785 ,clip,scale=1.,angle=0,keepaspectratio,width=0.81\textwidth]{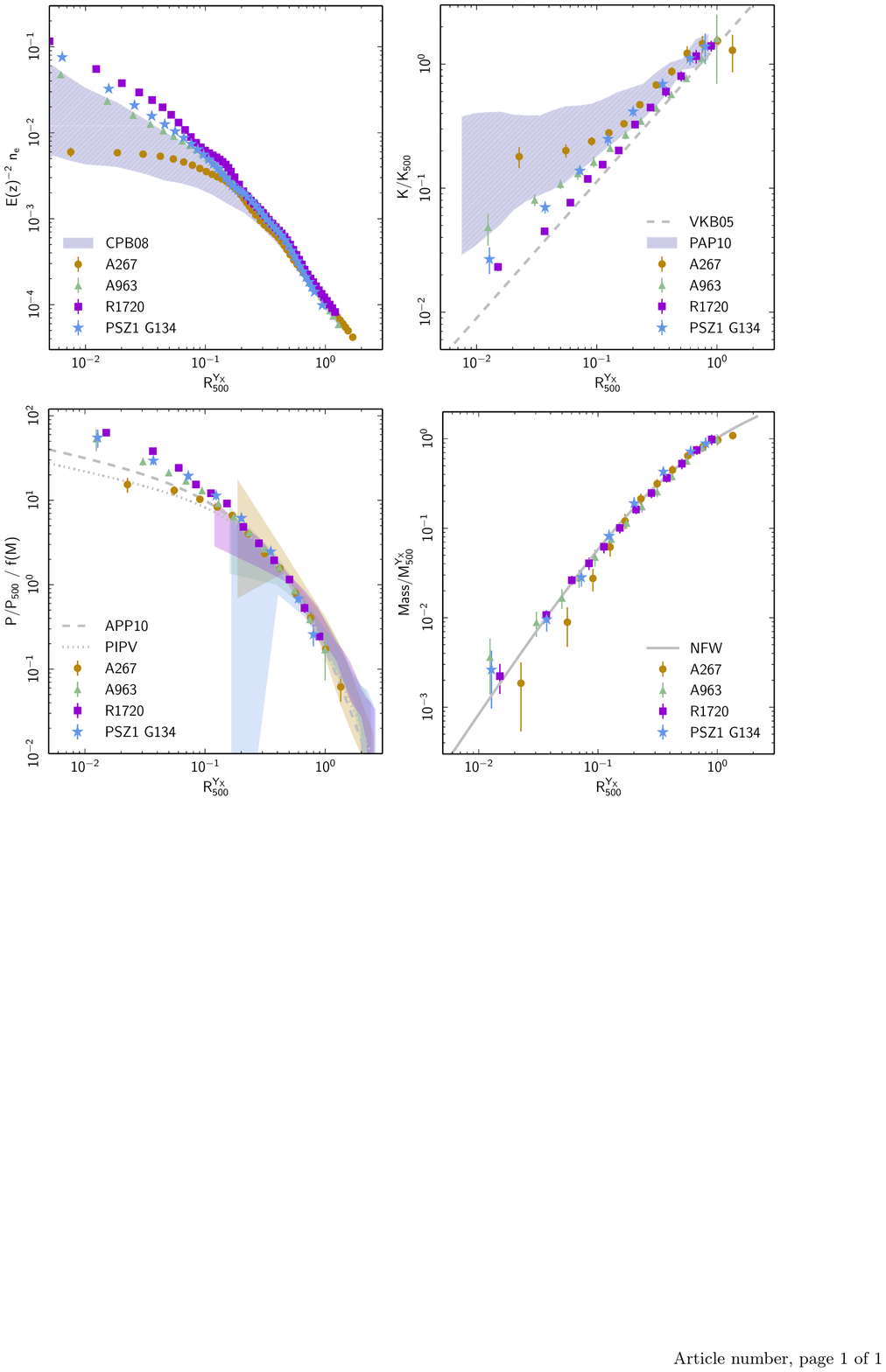}
\vspace{-0.7cm}
\end{center}
\caption{\footnotesize Top left: scaled density profiles compared to the $1\sigma$ dispersion of the \rexcess\ sample \citep{cro08}.  Top right: scaled entropy profiles compared to the $1\sigma$ dispersion of the \rexcess\ sample \citep{pra10}; the dashed line is the baseline gravitational entropy profile of \citet{vkb05}. Bottom left: scaled pressure profiles (points: X-ray, envelopes: SZ)  compared to the universal pressure profile of \citet{arn10} and the mean stacked profile of \citet{PIPV}. $f(M)$ is a (small) correction for the mass dependence in the pressure profile shape \citep[see][]{arn10}. Bottom right: scaled hydrostatic mass profiles compared to an NFW profile with $c=3.5$, for illustration only. Density, entropy, pressure, and mass profiles are renormalised by $E(z)^{-2}$, $K_{500}$, $P_{500}$, and $M_{500}^{Y_{\rm X}}$, respectively.
}
\label{fig:xrayprofs}
\end{figure*}
%-----------------------------Figure End--------------------------------

A key characteristic of the classification scheme advanced by \citet{jon03} is that the X-ray criterion is a lower limit. Indeed, a study of the Millennium Simulation has indicated that fossils may be found in significant numbers (3--4\% of the population) even in massive systems \citep{dar10}. Such high-mass systems are of  particular interest as they could potentially be the ultimate manifestation of gravitational collapse, while also being sites where non-gravitational effects are minimised. However, whether low- and high-mass fossil systems actually have a common origin is a subject of some debate (e.g., \citealt{dar10}, \citealt{har12}).

There are very few studies of high-mass fossil systems (although \citealt{sun04} and \citealt{kho06} have studied massive groups). The unique properties of the \planck\ SZ survey make it an ideal resource for finding such objects. The first \planck\ SZ catalogue (PSZ1, \citealt[][]{plaXXIX}) contains 813 confirmed clusters with known redshifts, of which 779 ($\sim 96$ \%) have $M_{500}^{Y_z} > 2 \times 10^{14}$ M$_{\odot}$\footnote{For this initial selection we use the SZ mass $M_{500}^{Y_z}$, calculated using the redshift and the SZ mass proxy $Y_z$, as described in PSZ1.}. Here we discuss the X-ray and SZ properties of four fossil clusters detected at high significance in the \planck\ PSZ1 survey. Combining \xmm\ X-ray and \planck\ SZ observations, we are able to detect the intra-cluster medium (ICM) out to well beyond $R_{500}$. Despite their very similar X-ray temperatures and total masses, we find a surprising variety in their hot gas content. We adopt a $\Lambda$CDM cosmology with $H_0=70$~km~s$^{-1}$~Mpc$^{-1}$, $\Omega_{\rm M}=0.3$, and $\Omega_\Lambda=0.7$; uncertainties are quoted at the 68\% confidence level.

%______________________________________________________________
% Various parameters
%
\begin{table*}[]
\caption[]{\footnotesize Sample data. }
\label{tab:sample}
\centering

\resizebox{\textwidth}{!} {
\begin{tabular}{@{}lccrccccrr@{}}

\toprule
\toprule
Cluster &  $z$   & $\Delta m_{12}$ & $T_{\rm X}$ & $L_{\rm X}$ & $M_{\rm 500, HE, NFW}$ & $c_{\rm 500, NFW}$ & $M_{500}^{\rm Y_X}$ & $f_{\rm gas, 500}$ & SNR \\
             &        &                              & (keV)       & ($10^{44}$ erg s$^{-1}$)   & ($10^{14}$ M$_{\odot}$) &   & ($10^{14}$ M$_{\odot}$) & &  \\

\midrule

\object{A267} & $ 0.227$ & $2.12$ & $5.48\pm0.15$ & $2.68\pm0.02$ & $3.60^{+0.26}_{-0.24}$ & $2.81\pm0.28$ & $3.73^{+0.07}_{-0.07} $ & $0.12\pm0.02$ & $4.54$ \\
\object{A963} & $ 0.206$ & $2.20$ & $5.59\pm0.12$ & $6.35\pm0.03$ & $4.82^{+0.58}_{-0.52}$ & $3.12\pm0.45$ & $5.02^{+0.07}_{-0.07}$ & $0.12\pm0.02$ & $7.96$ \\
\object{RX\,J1720.1+2638} & $0.164$ & $1.90$ & $5.80\pm0.12$ & $9.08\pm0.04$ & $5.26^{+0.63}_{-0.57}$ & $3.53\pm0.38$ & $5.30^{+0.08}_{-0.08}$ & $0.13\pm0.02$ & $10.75$ \\
\object{PSZ1\,G134.65$-$11.78} & 0.207 & 2.11 & $5.82\pm 0.23$ & $4.99\pm0.05$ & $5.78^{+0.98}_{-0.84}$ & $3.09\pm0.48$ & $4.34^{+0.17}_{-0.14}$ & $0.10\pm0.02^{*}$ & 4.80\\
\bottomrule
\end{tabular}
}
\tablefoot{Columns: (3): {\it r} -band magnitude gap between brightest- and second-brightest galaxies  (\citealt{har12,zar14}, Burenin, priv. comm.); (4): X-ray temperature in the $[0.15-0.75]\, \Rv$ region; (5) [$0.1-2.4]$\, keV X-ray luminosity; (6,7) mass and concentration from NFW fit; (8) mass from $M_{500}-Y_{\rm X}$ relation of \citet{arn10}; (9) gas mass fraction ($^{*}$ extrapolated from $0.94\,\Rv$); (10) PSZ1 SNR.\label{tab:pars}}
\end{table*}

%_______________

\section{The data}

We first cross-correlated the PSZ1 with three fossil catalogues: \citet{har12}, \citet{voe10}, and \citet{zar14}. Imposing a mass limit of $M_{500}^{Y_z} > 3.75 \times 10^{14}$ M$_{\odot}$ to select cluster-scale objects, we found three fossil clusters: A297, A963, and RXC\,J1720.1$+$2638. These are all very well-known systems that appear in \textit{ROSAT}-based X-ray catalogues, and they have all been observed in X-rays by \xmm. They are detected at signal-to-noise ratios ${\rm S/N > 4.5}$ in the \planck\ survey (see Table~\ref{tab:pars}).

We also examined data from the PSZ1 optical follow-up campaigns, in which a number of objects were identified as possible fossil systems \citep[see][]{RTT, ENO}. Applying the mass limit of $M_{500}^{Y_z} > 3.75 \times 10^{14}$ M$_{\odot}$ left us with two further candidates. During AO-14 we obtained \xmm\ observations of one of them, PSZ1\,G134.65$-$11.78, detected at a signal-to-noise ratio of ${\rm S/N}=4.8$ in the PSZ1. Only 25\% of the observation is usable, and the X-ray emission is traced out only to $0.94\, \Rv$, but we include it here for completeness.

The X-ray observations, data reduction, and procedures to derive the various radial profiles are described in detail in \citet{PC13}, where we investigated the scaling properties of these objects with their total SZ flux. Our X-ray data set consists of density, temperature, pressure, entropy, and hydrostatic mass profiles of each object. These last were fitted with an NFW profile to give an estimate of the concentration parameter $c_{500}$ and total mass $M_{\rm 500, HE, NFW}$; a second total mass estimate was obtained from iteration  about the $\Mv$--$Y_{\rm X}$ relation \citep{arn10}. 

For the SZ data, we used thermal SZ maps obtained from the full 30-month \planck\ data set \citep{YMAP} using the Modified Internal Linear Combination Algorithm \citep[MILCA;][]{hur13}. The SZ data reduction procedures are  described in detail in \citet{PIPV}, where we investigated the stacked pressure profile of a sample of local clusters. Our SZ data set consists of a pressure profile for each object. 

Table~\ref{tab:pars} lists various optical, X-ray, and SZ measurements for the four clusters, while Fig.~\ref{fig:xrayprofs} shows the radial profiles of density, entropy, pressure, and hydrostatic mass plotted on a logarithmic scale. Addition of the \planck\ data allows us to probe significantly further out in the cluster volume than is possible with the X-ray data alone, reconstructing the radial pressure distribution out to $R \sim 2.5\,\Rv$ in each case. 

 %-----------------------------Figure Start--------------------------------

\begin{figure}[]
\begin{center}
\includegraphics[scale=1.,angle=0,keepaspectratio,width=\columnwidth]{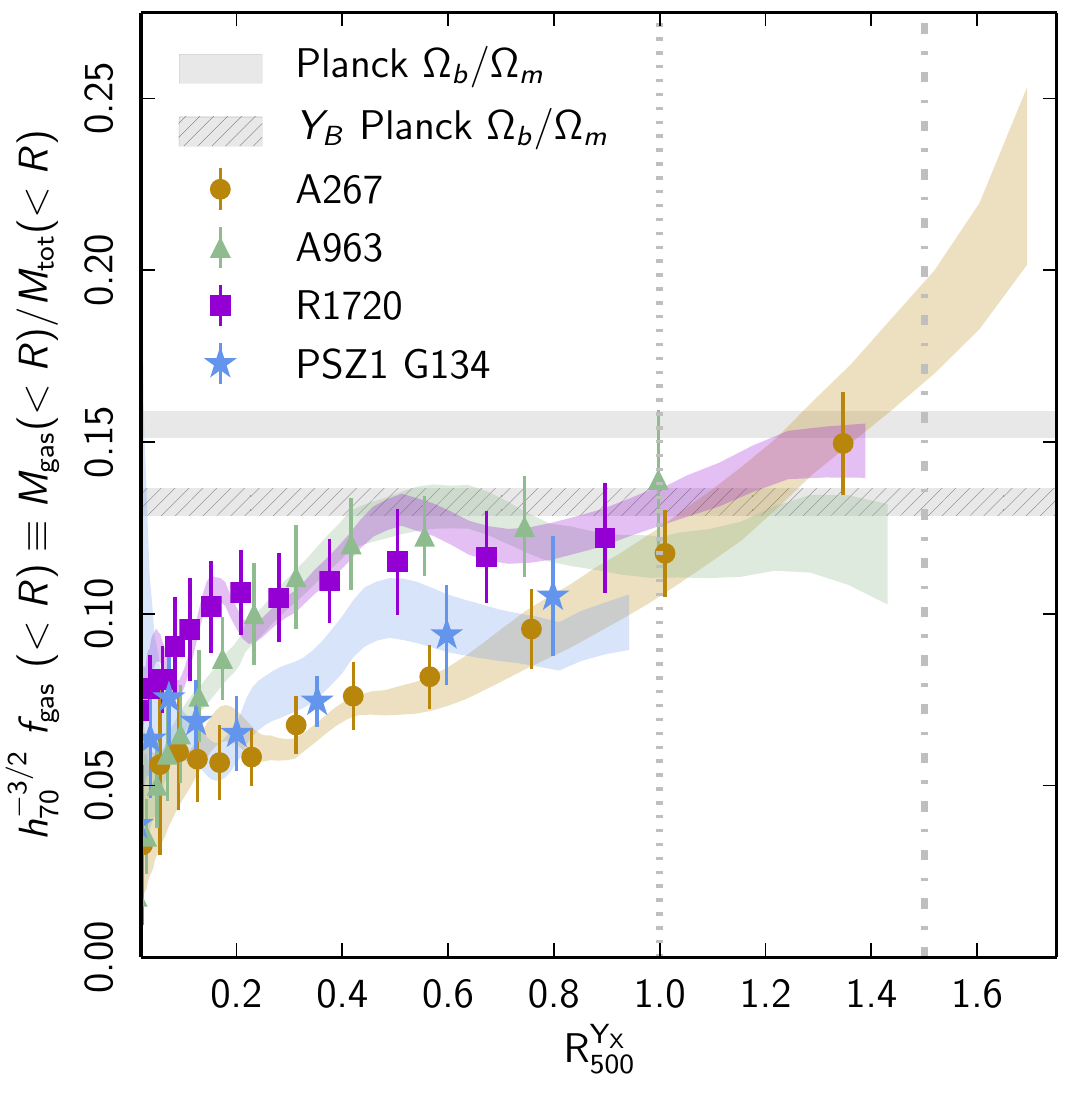}
\end{center}
\caption{\footnotesize Enclosed gas mass fraction profiles. Points with error bars indicate $f_{\rm gas}$ obtained from the X-day data only; coloured envelopes show the $f_{\rm gas}$ obtained from the X-ray-SZ pressure profile combined with the X-ray density profile (see Sect.~\ref{sec:fgas}). The solid grey line shows the cosmic baryon fraction measured by \citet{PXVI}; $Y_{\rm B} = 0.85$ is the baryon depletion factor at $\Rv$ from the simulations of \citet{pla13}. The dot-dashed line indicates $R_{200}\ (\sim 1.5\,\Rv$).
}
\label{fig:fgas}
\end{figure}
%-----------------------------Figure End--------------------------------

\section{General radial profile characteristics}

The total masses are confirmed to be high, and lie in a narrow range at $M_{500}^{Y_{\rm X}}  \sim 3.6-5.3 \times 10^{14}$ M$_{\odot}$, as expected from the small spread in temperatures ($5-6$ keV). For the three objects detected to at least $\Rv$, the total masses derived from the NFW fit to the integrated mass profile agree with those estimated from the $\Mv$--$Y_{\rm X}$ relation \citep{arn10} to better than $4\%$. This good agreement with a relation that was measured uniquely for relaxed systems would suggest that these fossil clusters may also be relaxed. The NFW concentration parameters also fall in a narrow range, and the mean concentration of $\langle c_{500} \rangle = 3.2\pm 0.4$ is the same as that measured for a sample of relaxed `normal' clusters by \citet{poi05}. These fossil clusters are thus not over-concentrated relative to normal relaxed systems. 

A general characteristic of the radial profiles shown in Fig.~\ref{fig:xrayprofs} is that three systems (A963, RX\,J1720.1+2638, and PSZ1\,G134.65$-$11.78) have remarkably similar profiles while those of A267 are clear  outliers, especially within $0.2\,\Rv$. These three fossil systems have the highly peaked central density and declining central temperature profiles  reminiscent of cool-core systems, while A267 has a flatter central density and temperature; indeed, its central density is an order of magnitude lower than that of RX\,J1720.1+2638. The entropy profiles converge towards the value expected from the non-radiative simulations of \citet{vkb05} at $\Rv$. 

In the central regions, the pressure profiles have somewhat less dispersion than the density, and beyond $R > 0.2\,\Rv$ they are remarkably similar; this is the case right out to the \planck\ detection limit of $R \sim 2.5\,\Rv$. The agreement between the X-ray and SZ pressure profiles is also good in the region of overlap. Beyond $\Rv$, the pressure profiles of all objects except A267 marginally exceed the mean universal profile of \citet{arn10} but agree with that of \citet{PIPV} at slightly more than $1\sigma$.

%%%%%%%%%%%%%%%%%%
\section{Gas mass fraction}\label{sec:fgas}

We took advantage of the extra radial reach afforded by the SZ observations to calculate the gas mass fraction by combining the SZ pressure and X-ray density profiles \citep[e.g.,][]{eck13}.
%\begin{equation}
%M_{\rm HE}\, (< R) = - \frac{R^2} {G \mu m_{\rm p} n_{\rm e}}  \frac{\dif P}{\dif R}; \label{eqn:eqnP}  
%\end{equation}
%\begin{equation}
%M_{\rm gas}\, (< R) = \mu m_{\rm p} \int^R_0 n_{\rm e} (R^\prime)\, 4 \pi R^{\prime 2} \dif R^{\prime}.
%\end{equation}
The gas pressure profile was obtained by fitting the X-ray and SZ data points (which together extend to $\sim 2.5\,\Rv$) with a generalised NFW model \citep{nag07}, with the uncertainties estimated using a Monte Carlo procedure. The gas mass profile was obtained from the X-ray density profiles, which extend to $0.94-1.7\,\Rv$ depending on the object. The gas mass fraction profiles $f_{\rm gas} (R) = M_{\rm gas}\, (< R) / M_{\rm HE}\, (< R)$ were then calculated in the radial range covered by the X-ray density profiles. In Fig.~\ref{fig:fgas} the results obtained from combining the X-ray and SZ information are compared to those from the X-ray data only,  plotted on a linear scale to accentuate the $R \gtrsim 0.2\,\Rv$ region. We also show the cosmic baryon fraction \citep{PXVI}, and the corresponding value corrected for the baryon depletion factor $Y_{B} = 0.85$ from the simulations of \citet{pla13}. 
 
These gas mass fraction profiles are striking. There is considerable diversity interior to $\Rv$; within  $R_{2500}$,  RX\,J1720.1+2638 and A963 have $30\%$ more gas than A267; indeed, they appear to have attained the depletion-corrected  cosmic baryon fraction by this radius, and flatten thereafter. The PSZ1\,G134.65$-$11.78 profile also seems to  flatten off at $R_{2500}$, but at a lower value of $f_{\rm gas}$; it resembles those of RX\,J1720.1+2638 and A963 but with lower normalisation. At $\Rv$ the median value of $f_{\rm gas,500} = 0.12\pm0.02$ agrees with recent observational \citep[e.g.,][]{lov15,ett15} and theoretical \citep[e.g.,][]{leb14} determinations, and with the depletion-corrected cosmic baryon fraction.

Beyond $\Rv$ the profiles diverge again; those of A963 and RX\,J1720.1+2638 remain flat, while that of A267 continues to increase out to the detection limit, reaching $f_{\rm gas} = 0.21\pm 0.01$ at $1.7\, \Rv$. This profile is highly reminiscent of that derived for Perseus by \citet{sim11}. As pointed out by these authors, 
if there is significant radially dependent clumping of the gas density, then the X-ray data will overestimate the gas content because the quantity we are measuring is the average of the square of the electron density $\langle n_{e}^2\rangle$, and not $\langle n_e \rangle$. An $f_{\rm gas}\sim 0.2$ would suggest an overestimate of the true gas density by a factor of $3-4$, depending on how much the gas content is affected by depletion. An alternative explanation for the high apparent $f_{\rm gas}$ is that it could be due to a departure from hydrostatic equilibrium (HE) in the outer regions, with the hydrostatic mass underestimating the true mass owing to neglect of non-thermal pressure support. That A267's gas mass fraction profile continues to climb while those of A963 and RX\,J1720.1+2638 remain flat thus suggests a considerable variation in gas clumping and/or HE beyond $\Rv$ in these fossil clusters.

%______________________________________________________________
% Dynamical indicators
\begin{table}[]
\caption{{\footnotesize Dynamical indicators.} 
\label{tab:dyn}} 
\centering  
\begin{tabular}{@{}lccc@{}}
\toprule
\toprule

\multicolumn{1}{c}{Cluster} & \multicolumn{1}{c}{Offset} & 
\multicolumn{1}{c}{$\langle w \rangle$} &
\multicolumn{1}{c}{$P_{\Delta}$} \\

\multicolumn{1}{c}{} & \multicolumn{1}{c}{(kpc)} & 
\multicolumn{1}{c}{} & \multicolumn{1}{c}{} \\

\midrule

A267 & 33 & $0.008$ & $< 0.01$ \\
A963 & 12 & $0.003$ & 0.86 \\
RX\,J1720.1+2638 & 2 & $0.004$ & 0.52 \\
PSZ1\,G134.65$-$11.78 & 12 & $0.006$ & \ldots \\

\bottomrule
\end{tabular}
\tablefoot{Columns: (2) offset between X-ray peak and BCG position; (3) standard deviation of the projected separations between the X-ray peak and centroid in ten equally-spaced radii in $[0.1-1]\ \Rv$; (4) Dressler-Shectman test probability. $P_{\Delta}<0.05$ is typical for disturbed systems \citep[e.g.,][]{pin96}.}
\end{table}
%______________________________________________________________

\section{Discussion and conclusions}

Combining data from \xmm\ and \planck, we have obtained the first high-quality measurements of the ICM of four bona fide fossil clusters out to unprecedented radii. Three systems have remarkably similar radial profiles reminiscent of cool-core systems. The fourth, A267, has the lowest inner gas content and shows the strongest evidence for clumping and/or departure from HE in the outskirts. It was classified as a fossil system by \citet{eig09} and \citet{zar14}. 

We undertook a number of X-ray and optical measurements to ascertain the dynamical status of our objects, summarised in Table~\ref{tab:dyn}. With the X-ray data we measured the offset between the X-ray peak and BCG position, and the centroid shift parameter $\langle w \rangle$ \cite[e.g.,][]{poo06,pra09}. In addition, we used the method proposed by \citet[][the DS test]{dre88} to probe for substructure in the galaxy distribution of A267, A963, and RX\,J1720.1+2638, for which extensive MMT/Hectospec and SDSS spectroscopic catalogues of $\sim$100-250 members per cluster are available \citep{rin13}. A267 has the highest centroid shift value, although it would be classified as unperturbed but non-cool-core according to the \rexcess\ criteria of \citet{pra09}. Its X-ray image is slightly elliptical and the offset between the position of its X-ray peak and BCG, at $\sim 33$ kpc, is more than three times that of the next most offset system. The DS test indicates that it is the only cluster that shows strong evidence of substructure in its galaxy distribution. 

If fossils truly represent old, evolved systems, then one would expect higher than average NFW concentrations owing to their early formation time. The average $c_{500}$ of our four fossil systems  is precisely the same as that obtained for `normal' local systems of a similar mass: they are not over-concentrated. 

Assuming similar formation scenarios for low- and high-mass fossils, our objects should have accreted the majority of their mass relatively early, and then remained undisturbed for a very long time. Growing today mainly through quiescent accretion, at fixed mass the hot gas in fossils would be in approximate HE, their radial profiles would be identical, and there would be small system-to-system variations in clumping factors and/or departures from HE in the outer regions. The surprising variety in the radial gas content of the systems studied here appears to be driven by variations in gas content interior to $\Rv$; conversely, outside $\Rv$ the dispersion may be driven by variations in clumping and/or HE.

Overall, these observations suggest that fossil systems cannot all be the ultimate stage of gravitational collapse, and that while some high-mass fossil objects appear to be dynamically active even today \citep[see also][]{zar16}, others are clearly relaxed but are not over-concentrated. Dedicated large-volume numerical simulations  will be needed to better clarify the formation history of the fossil population, especially at the highest masses that we have probed here for the first time.

\begin{acknowledgements}
We thank R. Burenin, R. Barrena, J. Démoclès and J.A. Rubi\~{n}o-Mart\'{i}n for their contributions to this project. This research has received funding from the European Research Council under the European Union’s Seventh Framework Programme (FP7/2007-2013)/ERC grant agreement no. 340519. EP acknowledges the support of the French Agence Nationale de la Recherche under grant ANR-11-BS56-015.
\end{acknowledgements}

\bibliographystyle{aa}
\bibliography{fossils}

\begin{thebibliography}{35}
\expandafter\ifx\csname natexlab\endcsname\relax\def\natexlab#1{#1}\fi

\bibitem[{{Arnaud} {et~al.}(2010){Arnaud}, {Pratt}, {Piffaretti},
  {B{\"o}hringer}, {Croston}, \& {Pointecouteau}}]{arn10}
{Arnaud}, M., {Pratt}, G.~W., {Piffaretti}, R., {et~al.} 2010, \aap, 517, A92

\bibitem[{Croston {et~al.}(2008)Croston, Pratt, B{\"o}hringer, Arnaud,
  Pointecouteau, Ponman, Sanderson, Temple, Bower, \& Donahue}]{cro08}
Croston, J.~H., Pratt, G.~W., B{\"o}hringer, H., {et~al.} 2008, \aap, 487, 431

\bibitem[{{Dariush} {et~al.}(2010){Dariush}, {Raychaudhury}, {Ponman},
  {Khosroshahi}, {Benson}, {Bower}, \& {Pearce}}]{dar10}
{Dariush}, A.~A., {Raychaudhury}, S., {Ponman}, T.~J., {et~al.} 2010, \mnras,
  405, 1873

\bibitem[{{Dressler} \& {Shectman}(1988)}]{dre88}
{Dressler}, A. \& {Shectman}, S.~A. 1988, \aj, 95, 985

\bibitem[{{Eckert} {et~al.}(2013){Eckert}, {Ettori}, {Molendi}, {Vazza}, \&
  {Paltani}}]{eck13}
{Eckert}, D., {Ettori}, S., {Molendi}, S., {Vazza}, F., \& {Paltani}, S. 2013,
  \aap, 551, A23

\bibitem[{{Eigenthaler} \& {Zeilinger}(2009)}]{eig09}
{Eigenthaler}, P. \& {Zeilinger}, W.~W. 2009, Astronomische Nachrichten, 330,
  978

\bibitem[{{Ettori}(2015)}]{ett15}
{Ettori}, S. 2015, \mnras, 446, 2629

\bibitem[{{Harrison} {et~al.}(2012){Harrison}, {Miller}, {Richards},
  {Lloyd-Davies}, {Hoyle}, {Romer}, {Mehrtens}, {Hilton}, {Stott}, {Capozzi},
  {Collins}, {Deadman}, {Liddle}, {Sahl{\'e}n}, {Stanford}, \& {Viana}}]{har12}
{Harrison}, C.~D., {Miller}, C.~J., {Richards}, J.~W., {et~al.} 2012, \apj,
  752, 12

\bibitem[{{Hurier} {et~al.}(2013){Hurier}, {Mac{\'{\i}}as-P{\'e}rez}, \&
  {Hildebrandt}}]{hur13}
{Hurier}, G., {Mac{\'{\i}}as-P{\'e}rez}, J.~F., \& {Hildebrandt}, S. 2013,
  \aap, 558, A118

\bibitem[{{Jones} {et~al.}(2003){Jones}, {Ponman}, {Horton}, {Babul},
  {Ebeling}, \& {Burke}}]{jon03}
{Jones}, L.~R., {Ponman}, T.~J., {Horton}, A., {et~al.} 2003, \mnras, 343, 627

\bibitem[{{Khosroshahi} {et~al.}(2006){Khosroshahi}, {Maughan}, {Ponman}, \&
  {Jones}}]{kho06}
{Khosroshahi}, H.~G., {Maughan}, B.~J., {Ponman}, T.~J., \& {Jones}, L.~R.
  2006, \mnras, 369, 1211

\bibitem[{{Le Brun} {et~al.}(2014){Le Brun}, {McCarthy}, {Schaye}, \&
  {Ponman}}]{leb14}
{Le Brun}, A.~M.~C., {McCarthy}, I.~G., {Schaye}, J., \& {Ponman}, T.~J. 2014,
  \mnras, 441, 1270

\bibitem[{{Lovisari} {et~al.}(2015){Lovisari}, {Reiprich}, \&
  {Schellenberger}}]{lov15}
{Lovisari}, L., {Reiprich}, T.~H., \& {Schellenberger}, G. 2015, \aap, 573,
  A118

\bibitem[{{Nagai} {et~al.}(2007){Nagai}, {Kravtsov}, \& {Vikhlinin}}]{nag07}
{Nagai}, D., {Kravtsov}, A.~V., \& {Vikhlinin}, A. 2007, \apj, 668, 1

\bibitem[{{Pinkney} {et~al.}(1996){Pinkney}, {Roettiger}, {Burns}, \&
  {Bird}}]{pin96}
{Pinkney}, J., {Roettiger}, K., {Burns}, J.~O., \& {Bird}, C.~M. 1996, \apjs,
  104, 1

\bibitem[{{Planck Collaboration Int. III}(2013)}]{PC13}
{Planck Collaboration Int. III}. 2013, \aap, 550, A129

\bibitem[{{Planck Collaboration Int. V}(2013)}]{PIPV}
{Planck Collaboration Int. V}. 2013, \aap, 550, A131

\bibitem[{{Planck Collaboration Int. XXVI}(2015)}]{RTT}
{Planck Collaboration Int. XXVI}. 2015, \aap, 582, A29

\bibitem[{{Planck Collaboration Int. XXXVI}(2016)}]{ENO}
{Planck Collaboration Int. XXXVI}. 2016, \aap, 586, A139

\bibitem[{{Planck Collaboration XVI}(2014)}]{PXVI}
{Planck Collaboration XVI}. 2014, \aap, 571, A16

\bibitem[{{Planck Collaboration XXII}(2015)}]{YMAP}
{Planck Collaboration XXII}. 2015, ArXiv e-prints [\eprint[arXiv]{1502.01596}]

\bibitem[{{Planck Collaboration XXIX}(2014)}]{plaXXIX}
{Planck Collaboration XXIX}. 2014, \aap, 571, A29

\bibitem[{{Planelles} {et~al.}(2013){Planelles}, {Borgani}, {Dolag}, {Ettori},
  {Fabjan}, {Murante}, \& {Tornatore}}]{pla13}
{Planelles}, S., {Borgani}, S., {Dolag}, K., {et~al.} 2013, \mnras, 431, 1487

\bibitem[{{Pointecouteau} {et~al.}(2005){Pointecouteau}, {Arnaud}, \&
  {Pratt}}]{poi05}
{Pointecouteau}, E., {Arnaud}, M., \& {Pratt}, G.~W. 2005, \aap, 435, 1

\bibitem[{{Ponman} {et~al.}(1994){Ponman}, {Allan}, {Jones}, {Merrifield},
  {McHardy}, {Lehto}, \& {Luppino}}]{pon94}
{Ponman}, T.~J., {Allan}, D.~J., {Jones}, L.~R., {et~al.} 1994, \nat, 369, 462

\bibitem[{{Poole} {et~al.}(2006){Poole}, {Fardal}, {Babul}, {McCarthy},
  {Quinn}, \& {Wadsley}}]{poo06}
{Poole}, G.~B., {Fardal}, M.~A., {Babul}, A., {et~al.} 2006, \mnras, 373, 881

\bibitem[{{Pratt} {et~al.}(2010){Pratt}, {Arnaud}, {Piffaretti},
  {B{\"o}hringer}, {Ponman}, {Croston}, {Voit}, {Borgani}, \& {Bower}}]{pra10}
{Pratt}, G.~W., {Arnaud}, M., {Piffaretti}, R., {et~al.} 2010, \aap, 511, A85

\bibitem[{{Pratt} {et~al.}(2009){Pratt}, {Croston}, {Arnaud}, \&
  {B{\"o}hringer}}]{pra09}
{Pratt}, G.~W., {Croston}, J.~H., {Arnaud}, M., \& {B{\"o}hringer}, H. 2009,
  \aap, 498, 361

\bibitem[{{Rines} {et~al.}(2013){Rines}, {Geller}, {Diaferio}, \&
  {Kurtz}}]{rin13}
{Rines}, K., {Geller}, M.~J., {Diaferio}, A., \& {Kurtz}, M.~J. 2013, \apj,
  767, 15

\bibitem[{{Simionescu} {et~al.}(2011){Simionescu}, {Allen}, {Mantz}, {Werner},
  {Takei}, {Morris}, {Fabian}, {Sanders}, {Nulsen}, {George}, \&
  {Taylor}}]{sim11}
{Simionescu}, A., {Allen}, S.~W., {Mantz}, A., {et~al.} 2011, Science, 331,
  1576

\bibitem[{{Sun} {et~al.}(2004){Sun}, {Forman}, {Vikhlinin}, {Hornstrup},
  {Jones}, \& {Murray}}]{sun04}
{Sun}, M., {Forman}, W., {Vikhlinin}, A., {et~al.} 2004, \apj, 612, 805

\bibitem[{{Voevodkin} {et~al.}(2010){Voevodkin}, {Borozdin}, {Heitmann},
  {Habib}, {Vikhlinin}, {Mescheryakov}, {Hornstrup}, \& {Burenin}}]{voe10}
{Voevodkin}, A., {Borozdin}, K., {Heitmann}, K., {et~al.} 2010, \apj, 708, 1376

\bibitem[{{Voit} {et~al.}(2005){Voit}, {Kay}, \& {Bryan}}]{vkb05}
{Voit}, G.~M., {Kay}, S.~T., \& {Bryan}, G.~L. 2005, \mnras, 364, 909

\bibitem[{{Zarattini} {et~al.}(2014){Zarattini}, {Barrena}, {Girardi},
  {Castro-Rodriguez}, {Boschin}, {Aguerri}, {M{\'e}ndez-Abreu},
  {S{\'a}nchez-Janssen}, {Catal{\'a}n-Torrecilla}, {Corsini}, {del Burgo},
  {D'Onghia}, {Herrera-Ruiz}, {Iglesias-P{\'a}ramo}, {Jimenez Bailon}, {Lozada
  Muoz}, {Napolitano}, \& {Vilchez}}]{zar14}
{Zarattini}, S., {Barrena}, R., {Girardi}, M., {et~al.} 2014, \aap, 565, A116

\bibitem[{{Zarattini} {et~al.}(2016){Zarattini}, {Girardi}, {Aguerri},
  {Boschin}, {Barrena}, {del Burgo}, {Castro-Rodriguez}, {Corsini}, {D'Onghia},
  {Kundert}, {M{\'e}ndez-Abreu}, \& {S{\'a}nchez-Janssen}}]{zar16}
{Zarattini}, S., {Girardi}, M., {Aguerri}, J.~A.~L., {et~al.} 2016, \aap, 586,
  A63

\end{thebibliography}

\raggedright
\end{document}